\documentclass[aps,prl,twocolumn,amsmath,amssymb]{revtex4}
\usepackage{graphicx}
\usepackage{amsmath}

\begin{document}

\title{Fermion Pairing across a Dipolar Interaction Induced Resonance}
\author{Ran Qi, Zhe-Yu Shi and Hui Zhai}
\affiliation{Institute for Advanced Study, Tsinghua University, Beijing, 100084, China}
\date{\today}
\begin{abstract}
It is known from the solution of the two-body problem that an anisotropic dipolar interaction can give rise to $s$-wave scattering resonances, which are named as dipolar interaction induced resonaces (DIIR). In this letter, we study zero-temperature many-body physics of a two-component Fermi gas across a DIIR. In the low-density regime, it is very striking that the resulting pairing order parameter is a nearly isotropic singlet pairing and the physics can be well described by an $s$-wave resonant interaction potential with finite range corrections, despite of the anisotropic nature of dipolar interaction. The pairing energy is as strong as a unitary Fermi gas nearby a magnetic Feshbach resonance. In the high density regime, the anisotropic effect plays an important role. We find phase transitions from singlet pairing to a state with mixed singlet and triplet pairing, and then from mixed pairing to pure triplet pairing. The state with mixed pairing spontaneously breaks the time-reversal symmetry.

\end{abstract}
\maketitle

With the development of STIRAP technique \cite{STIRAP}, it is now very promising to achieve degenerate gases of polar molecules within few years. The dipole moment of a polar molecule can be tuned continuously  by increasing the strength of applied external electric field. Because of the large permanent dipole moment, the dipolar interaction energy can be on the same order of magnitude as the Fermi energy. Previously, many studies of dipolar Fermi gases focus on $p$-wave pairing due to the anisotropic nature of dipolar interactions \cite{p-wave,p-wave2,cooper}. However, the transition temperature for $p$-wave superfluid is usually suppressed by the centrifugal barrier. Thus, if the strong dipolar interaction can also cause significant effects in the $s$-wave channel, it must be first seen in experiments once a chemically stable degenerate polar molecular gas is realized.

The interaction potential between two molecules with dipole moment $d\hat{z}$ is given by (we also set $\hbar=k_B=1$ throughout this letter):
\begin{equation}
V_{\text{d}}({\bf r})=\frac{2D}{m}\frac{1-3\cos^2\theta_{\mathbf{r}}}{r^3} \label{Vd0}
\end{equation}
where $D=md^2/2$ is dipole length, $\theta_{{\bf r}}$ is the polar angle between ${\bf r}$ and $\hat{z}$. At first, one can easily see that $\langle s|V_{\text{d}}({\bf r})|s\rangle=0$ for any $s$-wave wave-function $|s\rangle$. Thus, within Born approximation the interaction effect in the $s$-wave channel is always weak. However, because $V_{\text{d}}({\bf r})$ couples s-wave $|s\rangle$ to d-wave $|d\rangle$, and through a second-order process, it induces an effective attractive potential proportional to $-D^2/r^4$ in the $s$-wave channel \cite{note}. Thus, as $D$ increases, the potential becomes deeper and deeper, and therefore $s$-wave bound states will appear sequently. This will lead to a series of scattering resonances in the $s$-wave channel \cite{DIIR_You,DIIR_Blume,DIIR,John,Shi}, which are named as dipolar interaction induced resonances (DIIR). Two-body solution shows that nearby an $s$-wave DIIR, the scattering amplitude is dominated by s-wave component and is very isotropic for low-energy scattering  \cite{DIIR}, and the scattering phase shift is strongly momentum dependent, which gives rise to  a large and positive effective range proportional to $D$ \cite{Shi}. This is strongly in contrast to a wide Feshbach resonance studied extensively before, where the effective range is negligible.

One major difficulty in studying many-body physics of dipolar gases is that the bare interaction Eq. (\ref{Vd0}) has short range divergency.
This is because the Fourier transform  $V_{\text{d}}({\bf k})$ of $V_{\text{d}}({\bf r})$ approaches a constant as $k\rightarrow \infty$, thus, it has the similar ultraviolet divergency as a $\delta$ function contact potential.
One common approach to eliminate this divergency is to relate $V_{\text{d}}(\mathbf{k-k'})$ to scattering amplitude $\Gamma_{\mathbf{k,k'}}$ via Lippman-Schwinger equation as
\begin{equation}
\Gamma_{\mathbf{k,k'}}=V_{\text{d}}(\mathbf{k-k'})-\sum_{\mathbf{q}}\Gamma_{\mathbf{k,q}}\frac{m}{q^2}V_{\text{d}}(\mathbf{q-k'}).
\end{equation}
This approach is widely used in regularizing a $\delta$-function interaction potential \cite{crossover}. Nevertheless, for anisotropic dipolar interaction potential, such an equation is difficult to solve. So far most of previous studies of many-body physics of dipolar Fermi gas are based on the Born approximation, which assumes $\Gamma_{\mathbf{k,k'}}=V_{\text{d}}(\mathbf{k-k'})$. It greatly simplifies the solution, but the resulting $\Gamma_{\mathbf{k,k'}}$ is always anisotropic and clearly it fails near a DIIR. On the other hand, so far the studies of DIIR are limited to either two-body physics \cite{DIIR_You,DIIR_Blume,DIIR,John,Shi} or the high-temperature regime \cite{Shi}.

In this letter we study the zero temperature many-body physics for a two-component Fermi gas across a DIIR. Here we choose an alternative way to overcome the regularization problem, that is, by introducing a regular model potential at short distance, which is given by \begin{eqnarray}
V(\mathbf{r})=\frac{2D}{m}\frac{3\cos^2\theta_{\mathbf{r}}-1}{r^3}F(r/r_0)\label{Vd}
\end{eqnarray}
where $F(x)$ is chosen as $F(x)=e^{-x}(x^3/6+x^2/2+x+1)-1$. We choose this particular form of model potential because its Fourier transform $V({\bf k})$ has an analytical expression
\begin{equation}
V(\mathbf{k})=\frac{8\pi D}{3m}\frac{3\cos^2\theta_{\mathbf{k}}-1}{(1+k^2r_0^2)^2}
\end{equation}
As shown in Fig. 1, $F(x)\rightarrow -1$ exponentially as $x\gg 1$, thus $V({\bf r})$ behaves the same as the bare dipole interaction $V_\text{d}({\bf r})$ when $r\gg r_0$, and $F(x)\rightarrow x^4$ as $x\rightarrow 0$, which ensures $V({\bf k})$ decays sufficiently fast as $k\rightarrow\infty$. More importantly, we should note that only the physics nearby a DIIR is independent of how we implement the regularization. Thus, we need to first locate the DIIR for this particular potential by solving the two-body Scr$\ddot{o}$dinger equation. Indeed, we find the first DIIR at $D/r_0=8.81$ with a positive effective range $r_{e}=1.15D$, as shown in Fig. 1(b) and (d). Our following many-body study will be restricted to the regime nearby this resonance.

{\it Mean-field Theory.} The many-body Hamiltonian for this two-component Fermi gas is given by
\begin{eqnarray}
\!\hat{H}\!=\!\sum_{\alpha\mathbf{k}}\!\epsilon_{\mathbf{k}}\hat{\psi}^{\dag}_{\alpha\mathbf{k}}\hat{\psi}_{\alpha\mathbf{k}}\!+\!
\frac{1}{2V}\!\sum_{\mathbf{kk'q}}^{\alpha\beta}\hat{\Pi}^{\dag}_{\alpha\beta}(\mathbf{k\!,\!q})\!V(\mathbf{k\!-\!k'})\!\hat{\Pi}_{\alpha\beta}(\mathbf{k'\!,\!q})\label{H}
\end{eqnarray}
where $\epsilon_{\mathbf{k}}={\bf k}^2/(2m)$, $\alpha$ and $\beta$ denote spin $\uparrow$ and $\downarrow$, and $\hat{\Pi}_{\alpha\beta}(\mathbf{k,q})=\hat{\psi}_{\alpha\mathbf{q/2-k}}\hat{\psi}_{\beta\mathbf{q/2+k}}$. Here the spin refers to hyperfine spin degree of a polar molecule \cite{hyperfine}, while the dipolar interaction is originated from the electronic dipole moment, thus we can take the dipole interaction as spin independent. Therefore, this Hamiltonian possesses an $SU(2)$ spin rotational symmetry. Following the standard mean-field approach, we decouple the interaction term in the Cooper channel and obtain
\begin{eqnarray}
\!\hat{H}_{\text{mf}}&\!=\!&\sum_{\alpha\mathbf{k}}\!\epsilon_{\mathbf{k}}\hat{\psi}^{\dag}_{\alpha\mathbf{k}}\hat{\psi}_{\alpha\mathbf{k}}\!+\!
\frac{1}{2V}\!\sum_{\alpha\beta\mathbf{k}}[\hat{\Pi}^{\dag}_{\alpha\beta}(\mathbf{k\!,0})\Delta_{\alpha\beta}(\mathbf{k})\!+\!\text{h.c.}]\nonumber\\
&\!-\!&\frac{1}{2}\sum_{\alpha\beta\mathbf{k,k'}}\Delta^*_{\alpha\beta}(\mathbf{k})V^{-1}(\mathbf{k,k'})\Delta_{\alpha\beta}(\mathbf{k'})\label{Hmf}
\end{eqnarray}
where $\Delta_{\alpha\beta}(\mathbf{k})=\sum_{{\bf k'}} V(\mathbf{k\!-\!k'})\langle\hat{\Pi}_{\alpha\beta}({\bf k'},0)\rangle$ and $V^{-1}(\mathbf{k,k'})$ satisfies $\sum_{\mathbf{k''}}V^{-1}(\mathbf{k,k''})V(\mathbf{k''-k'})=\delta_{\mathbf{k,k'}}$. The four superfluid order parameters $\Delta_{\alpha\beta}$ can be organized into one singlet and three triplet components as $\Delta_{\text{s}}=\frac{1}{\sqrt{2}}(\Delta_{\uparrow\downarrow}-\Delta_{\downarrow\uparrow})$,
$\Delta_{\text{t}}^z=\frac{1}{\sqrt{2}}(\Delta_{\uparrow\downarrow}+\Delta_{\downarrow\uparrow})$, $\Delta_{\text{t}}^x=-\frac{1}{\sqrt{2}}(\Delta_{\uparrow\uparrow}-\Delta_{\downarrow\downarrow})$ and
$\Delta_{\text{t}}^y=-\frac{i}{\sqrt{2}}(\Delta_{\uparrow\uparrow}+\Delta_{\downarrow\downarrow})$.

\begin{figure}[t]
\includegraphics[height=2.3 in, width=3.1 in]
{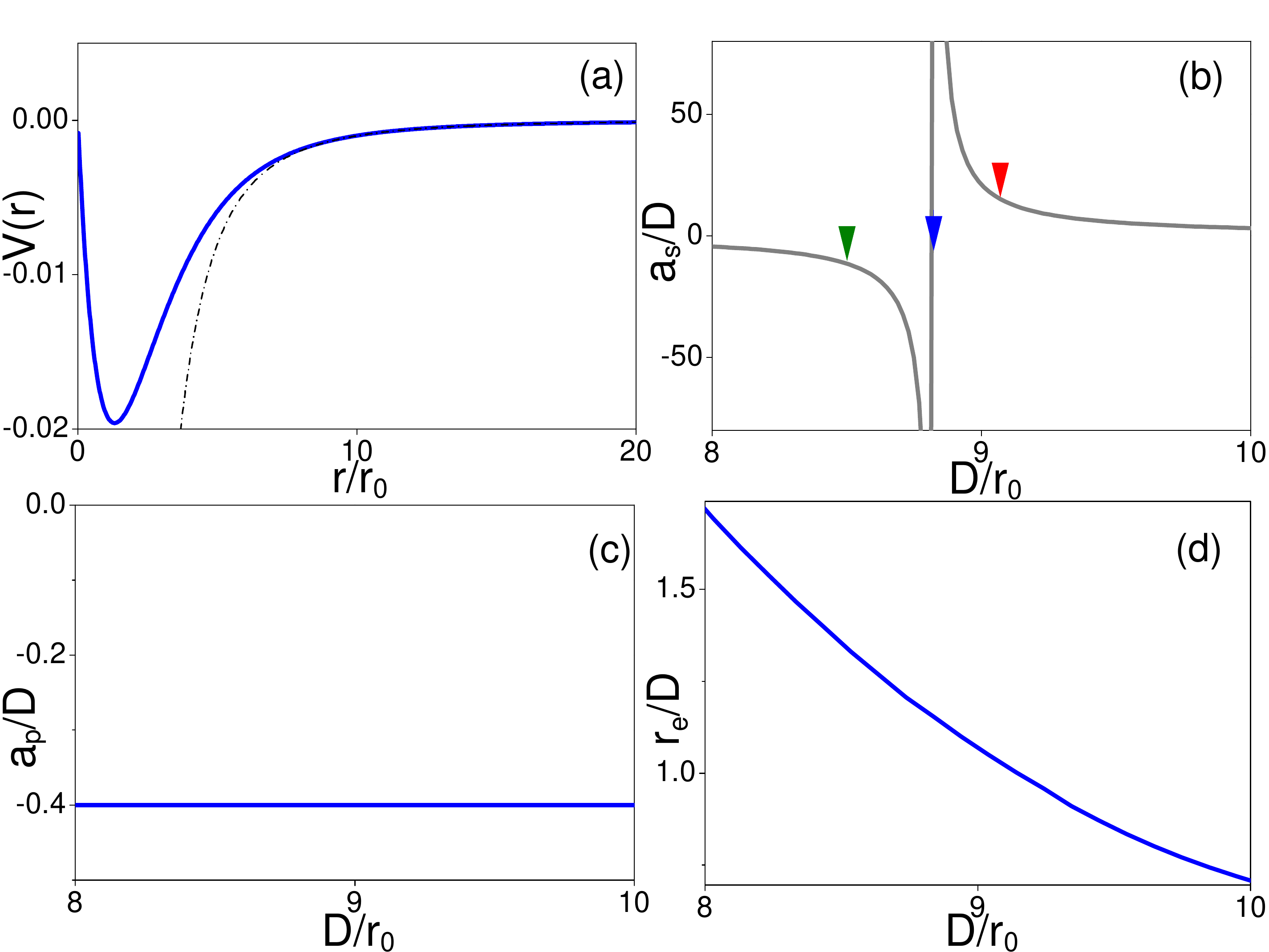}
\caption{(a) Regularized dipolar interaction potential $V({\bf r})$, and the dashed line indicates the long range dipolar part. (b-d) $s$-wave scattering length $a_{\text{s}}$ (b) $p$-wave scattering length $a_{\text{p}}$ (c) and the $s$-wave effective range $r_{\text{e}}$ (d) around a DIIR. The plot is around the first DIIR  located at $D/r_0=8.81$.\label{as}}
\end{figure}

Tracing out the fermion part, it yields the free energy:
\begin{align}
&\Omega_0\!=\!\! \frac{1}{2}\sum_{\mathbf{k},j=1,2}\Big[\xi_{\mathbf{k}}-2T\ln\big(2\cosh \frac{E_{j\mathbf{k}}}{2T} \big) \Big]
\nonumber\\
&\!-\! \frac{1}{2}\!\sum_{\mathbf{k,k'}}\!\Big[\Delta^*_{s\mathbf{k}}V^{-1}_{s}(\mathbf{k,k'})\Delta_{s\mathbf{k'}} \!+\!\sum_{\nu=x,y,z}\Delta^{\nu*}_{t\mathbf{k}}V^{-1}_{t}(\mathbf{k,k'})\Delta^{\nu}_{t\mathbf{k'}}\Big]
\label{Omega}
\end{align}
where $\xi_{\mathbf{k}}=\epsilon_{\mathbf{k}}-\mu$, $V_{s,t}(\mathbf{k,k'})=[V(\mathbf{k+k'})\pm V(\mathbf{k-k'})]/2$,
 $E_{j\mathbf{k}}=\sqrt{\xi_{\mathbf{k}}^2+\lambda_{j\mathbf{k}}^2}$, and $\lambda_{1,2\mathbf{k}}^2=[L_1\pm\sqrt{L_1^2-L_2^2}]/2$
with $ L_1=\sum\limits_{\nu=x,y,z}|\Delta^{\nu}_{t\mathbf{k}}|^2+|\Delta_{s\mathbf{k}}|^2$ and $L_2=\big|\sum\limits_{\nu=x,y,z}(\Delta^{\nu}_{t\mathbf{k}})^2-(\Delta_{s\mathbf{k}})^2\big|$. Both $L_1$ and $L_2$ are invariant under an $SU(2)$ spin rotation. Therefore, in analogy with spin-1 Bose condensate \cite{spin1}, we can classify the triplet components into two different types similar as polar phase (with $\Delta_{t\mathbf{k}}^z=\Delta_{t\mathbf{k}},~\Delta_{t\mathbf{k}}^x=\Delta_{t\mathbf{k}}^y=0$) and ferromagnetic phase (with $\Delta_{t\mathbf{k}}^z=0,~\Delta_{t\mathbf{k}}^x=i\Delta_{t\mathbf{k}}^y=\Delta_{t\mathbf{k}}$). It is easy to show that the ferromagnetic configuration is always energetically unfavored and will introduce a finite magnetization  \cite{SU2}, while in this work we are interested in equal population case. Hereafter we will only consider the polar type configuration. One can further show that the free energy is minimized when $\Delta_{s\mathbf{k}}$ and $\Delta_{t\mathbf{k}}$ have a relative phase $\pi/2$ which leads to
 $\lambda_{1,2\mathbf{k}}^2=\lambda_{\mathbf{k}}^2=(|\Delta_{s\mathbf{k}}|^2+|\Delta_{t\mathbf{k}}|^2)/2$ and $E_{1,2\mathbf{k}}=E_{\mathbf{k}}=\sqrt{\xi_{\mathbf{k}}^2+\lambda_{\mathbf{k}}^2}$ \cite{congjun}.

Finally, we obtain the gap equation for $\Delta_{s\mathbf{k}}$ and $\Delta_{t\mathbf{k}}$ by minimizing the free energy functional Eq. (\ref{Omega}) as
\begin{eqnarray}
\Delta_{\zeta\mathbf{k}}=-\int\frac{d^3 \mathbf{q}}{(2\pi)^3}V_{\zeta}(\mathbf{k,q})\frac{\tanh\frac{E_{\mathbf{q}}}{2T}}{2E_{\mathbf{q}}}\Delta_{\zeta\mathbf{q}}\label{gap}
\end{eqnarray}
where $\zeta=s,t$, respectively. We note that in Eqs. \ref{gap}, $\Delta_s$ and $\Delta_t$ are not decoupled. In fact, they are coupled through $E_{{\bf q}}$ which depends on both $\Delta_s$ and $\Delta_t$. The gap equation (\ref{gap}) can be solved numerically together with the number equation $n=-\partial\Omega_0/\partial\mu=\int d^3\mathbf{k}n_{\mathbf{k}}/(2\pi)^3$ where
 $n_{\mathbf{k}}=1-\frac{\xi_{\mathbf{k}}}{E_{\mathbf{k}}}\tanh\frac{E_{\mathbf{k}}}{2T}$ and $n=n_{\uparrow}+n_{\downarrow}=k_F^3/(3\pi^2)$ is the total density, with $k_\text{F}$ the Fermi momentum of each spin component. Due to the azimuthal symmetry of the interaction potential, we shall consider the solution that $\Delta_{\zeta\mathbf{k}}$ only depends on $k=|{\bf k}|$ and $\cos\theta_{\mathbf{k}}$, and is independent of the azimuthal angle $\phi_{\mathbf{k}}$ of ${\bf k}$. Furthermore, we can expand $\Delta_{\zeta{\mathbf k}}$ into different partial waves as $\Delta_{\zeta{\mathbf k}}=\sum_{\ell}\Delta_{\zeta\ell}(k)P_{\ell}(\cos\theta_{\mathbf{k}})$ where $P_{\ell}(x)$ is the Legrend polynomials and $\ell$ runs over even/odd integers for $\zeta=s/t$.

{\it Results:} In this problem, we can choose two dimensionless parameters $D/a_{\text{s}}$ and $k_{\text{F}}D$ as control parameters of the many-body physics. Since we focus on the regime nearby the first DIIR, there is one-to-one correspondence between $D/a_{\text{s}}$ and $D/r_0$ as shown in Fig. 1(b). And once $D/r_0$ is fixed, the other parameters of low-energy scattering process, such as $s$-wave effective range $r_e/D$ and $p$-wave scattering length $a_{\text{p}}/D$ are also fixed, as shown in Fig. 1(c) and (d). Another parameter $k_\text{F}D$ controls the ratio of dipole length $D$, as well as the effective range $r_e$, to the average inter-particle distance. The main results of this work are:

(A) Nearby resonance where $D/a_{\text{s}}=0$, for small $k_{\text{F}}D$, the system is in a pure singlet pairing phase. In this regime, despite that the dipolar interaction is anisotropic, the resulting order parameter is nearly isotropic. Moreover, we show that in this regime the physics can be well captured by a simpler model, that is, a pure isotropic $s$-wave resonant interaction with a finite range correction.

(B) As $k_{\text{F}}D$ increases, the system first undergoes a transition to a mixed parity phase with both non-zero singlet and triplet order parameters, which spontaneously breaks the time-reversal symmetry. When $k_{\text{F}}D$ further increases, it experiences another phase transition to a pure triplet phase. A phase diagram across DIIR in terms of $D/a_{\text{s}}$ and $k_{\text{F}}D$ is constructed.

\begin{figure}[t]
\includegraphics[height=2.7 in, width=3.2 in]
{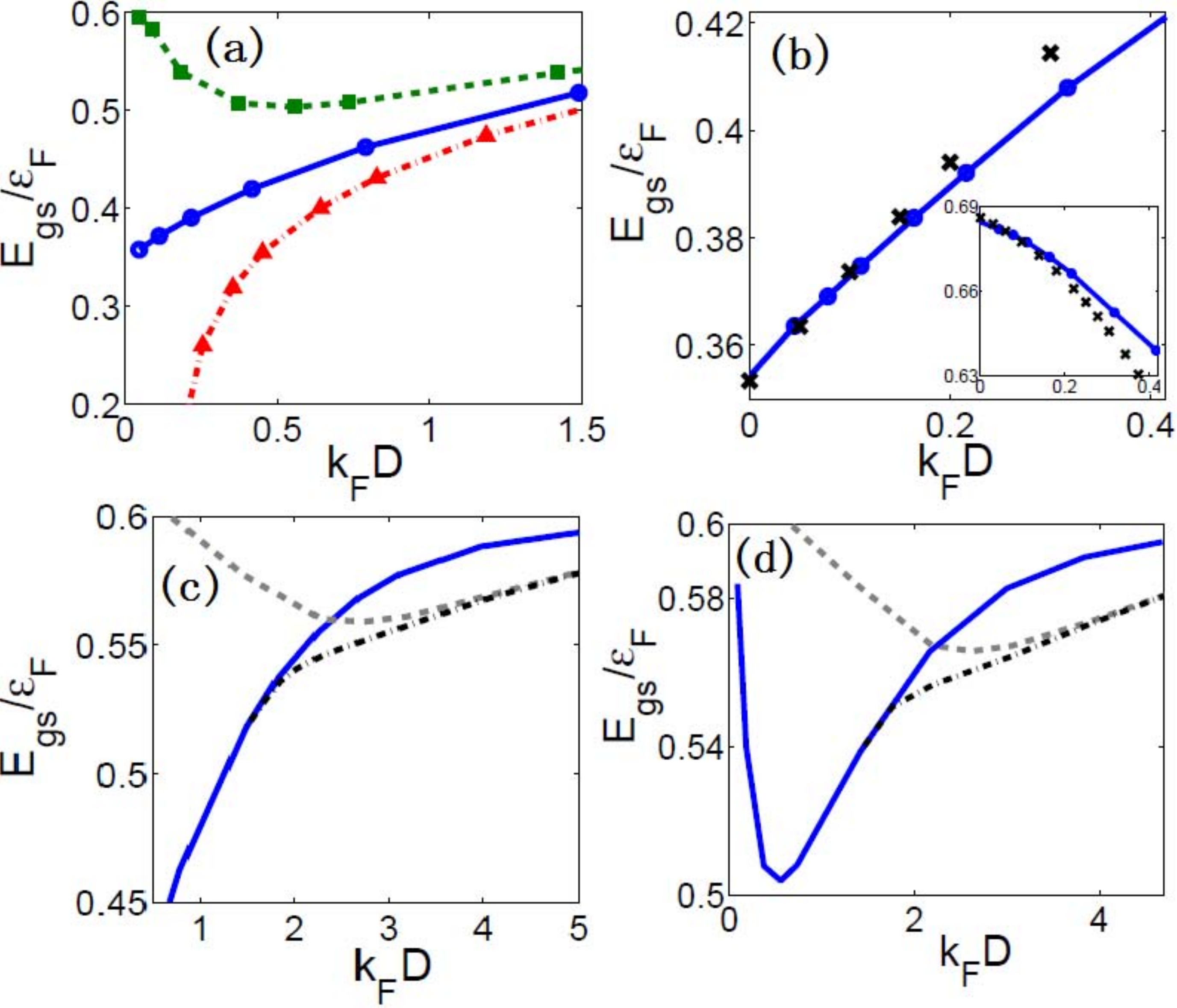}
\caption{(a) Ground state energy per particle as a function of $k_\text{F}D$ at fixed $D/r_0$ marked in Fig. \ref{as}(b), $D/a_{\text{s}}<0$ for dash green line with squares, $D/a_{\text{s}}=0$ for blue solid line with circles and $D/a_{\text{s}}>0$ for dash dotted red line with triangles. (b) Comparison of ground state energy between dipolar problem and $s$-wave separable potential calculation (black crosses) with finite effective range. The inset: the comparison of order parameter. (c-d) The energy of pure singlet pairing (solid line), pure triplet pairing (dashed line) and mixed pairing (dash-dotted line). (c) is at the resonance point and (d) is at the BCS side.  \label{energy}}
\end{figure}

Hereafter we shall explain these two results in details.

{\it (A) Small $k_{\text{F}}D$ Regime:} We find that for small $k_{\text{F}}D$ the solution of gap equation Eqs. \ref{gap} gives pure singlet pairing, i.e. $\Delta_{s\mathbf{k}}\neq 0$ and $\Delta_{t\mathbf{k}}=0$. In Fig. \ref{energy}(a) we plot the ground state energy per particle $E_{\text{gs}}$ as a function of $k_\text{F}D$ for three different values of $D/r_0$, as marked in Fig. 1(b). As noted above, once $D/r_0$ is fixed, $D/a_{\text{s}}$ is also fixed, and the three curves in Fig. \ref{energy}(a) correspond to $D/a_{\text{s}}<0$, $=0$ and $>0$, respectively. If we take the limit $k_{\text{F}}D\rightarrow 0$, $k_\text{F}a_\text{s}\rightarrow 0^-$ or $0^+$ for $D/a_{\text{s}}<0$ and $>0$ case, respectively. As shown in Fig.  \ref{energy}(a), $E_{\text{gs}}$ approaches free fermion case of $0.6E_{\text{F}}$ in $k_\text{F}a_\text{s}\rightarrow 0^-$ case, while $E_{\text{gs}}$ continuously decreases in $k_\text{F}a_\text{s}\rightarrow 0^+$ case because all fermions form deep bound states. If $D/a_{\text{s}}=0$, as $k_\text{F}D\rightarrow 0$, $E_{\text{gs}}$ approaches $0.353E_\text{F}$, and very surprisingly, this coincides with the mean-field results of unitary Fermi gases with single channel $s$-wave contact potential. That is to say, in the regime of small $k_{\text{F}}D$ and nearby a DIIR, the pairing energy is on the same order of the Fermi energy as in a unitary Fermi gas at a magnetic Feshbach resonance, and thus, its transition temperature is expected to be as high as a unitary Fermi gas.

\begin{figure}[t]
\includegraphics[height=2.5 in, width=3.3 in]
{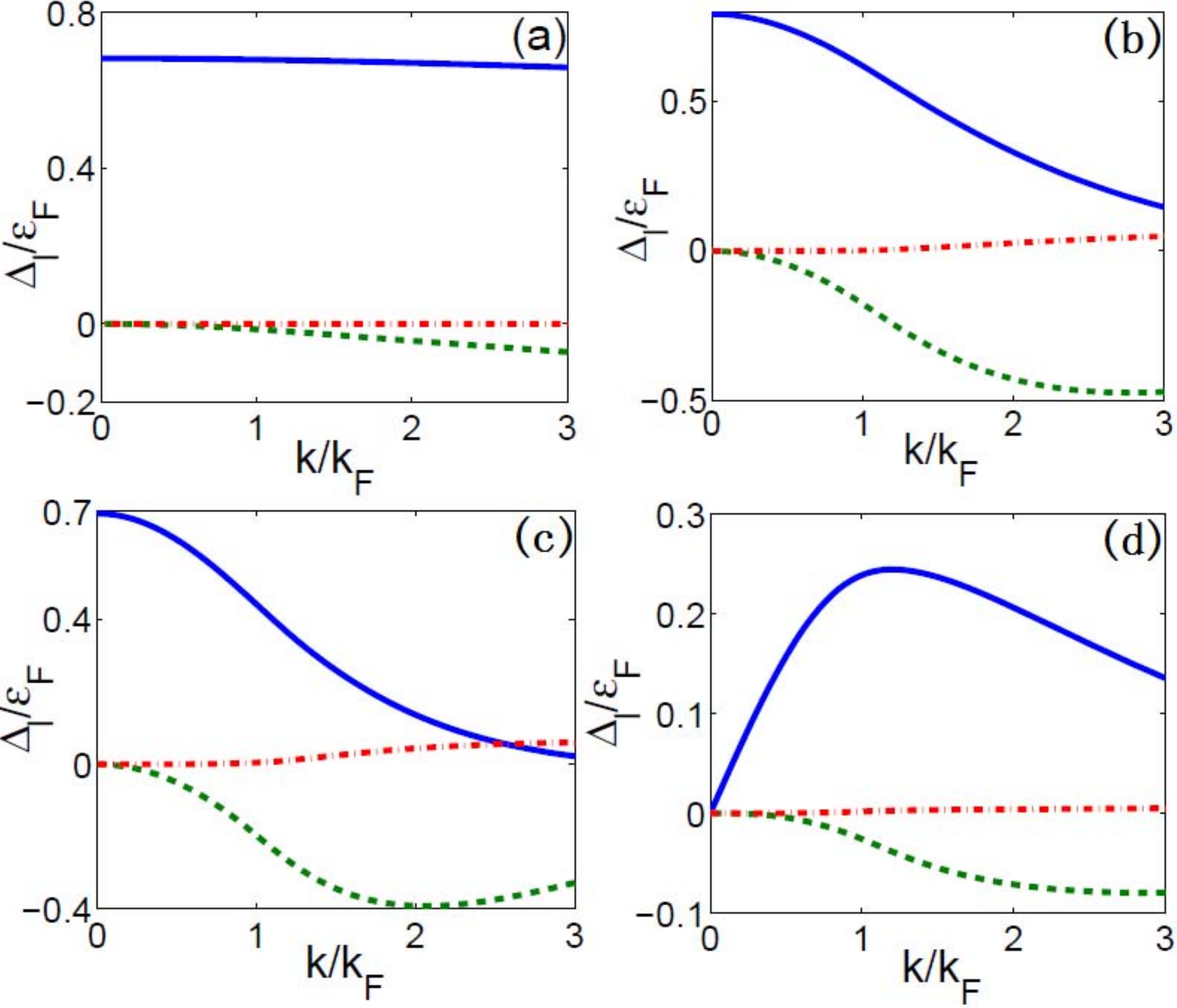}
\caption{Partial wave distribution of superfluid order parameter right at the first DIIR. (a)-(c) pure singlet pairing with $k_\text{F}D=0.1$, $1.5$ and $2.2$ for (a), (b) and (c), respectively. $\ell=0$ for blue solid line; $\ell=2$ for green dashed line; $\ell=4$ for red dash-dotted line. (d) pure triplet pairing with $k_\text{F}D=1.5$: $\ell=$1, 3, 5 corresponding to the same symbols as $\ell=$0, 2, 4 in (a-c). \label{Deltal}}
\end{figure}

Such behaviors are strongly in contrast to what one can obtain from Born approximation, where the singlet pairing is always exponentially small as $k_\text{F}D\rightarrow 0$. Thus, these are unique features due to scattering resonances. Furthermore, although the dipolar potential is anisotropic, the solution of order parameter is very isotropic in this regime. As shown in Fig. \ref{Deltal}(a), the order parameter is dominated by $s$-wave, and the contribution from $d$-wave and even higher waves are negligibly small. This is in fact consistent with the physical insight from two-body problem, where the scattering amplitude becomes very isotropic at DIIR and the contributions of high partial waves become very small \cite{DIIR,Shi}.

Based on these observations, we can further show that the physics of this regime can be captured by a simpler model of an s-wave separable potential  $V_{\mathbf{kk'}}^{s}=V_0w_kw_{k'}$, where $w_{k}=\Theta(\Lambda-k)$ and a momentum cutoff $\Lambda$ is introduced to implement a finite effective range. By solving two-body problem with $V_{\mathbf{kk'}}^{s}$ one finds $a_{\text{s}}=\infty$ by fixing $mV_0\Lambda=-2\pi^2$, which possess a finite effective range $r_e=4/(\pi\Lambda)$. Thus, we fix $k_{\text{F}}/\Lambda=1.15\pi k_{\text{F}}D/4$ so that the effective range from this s-wave separable model is fixed to the same value as the dipolar potential.
As shown in Fig. \ref{energy}(b), when $k_{\text{F}}D\lesssim 0.3$, the ground state energy and the strength of order parameters obtained from this $s$-wave model agree very well with dipolar model at DIIR.

The finite range correction can explain why energy increases as $k_{\text{F}}D$ increases in this regime. Because we can write an energy dependent scattering length $a_{\text{s}}(k)$ as $1/a_{\text{s}}(k)=1/a_{\text{s}}-r_e k^2/2$, thus, with positive $r_e$, $a_{\text{s}}(k)$ is negative at DIIR for finite $k$. At the Fermi energy, $1/(k_{\text{F}}a_{\text{s}}(k_{\text{F}}))=1/(k_{\text{F}}a_{\text{s}})-k_{\text{F}}r_e/2$. Thus, as $k_{\text{F}}D$ increases, $k_{\text{F}}r_e$ increases and the attraction around the Fermi surface becomes weaker. Thus, the energy increases. Recently, several other papers have also studied finite range correction to an $s$-wave model \cite{effectiverange} and our results are consistent with theirs.

\begin{figure}[t]
\includegraphics[height=2.3 in, width=3.2 in]{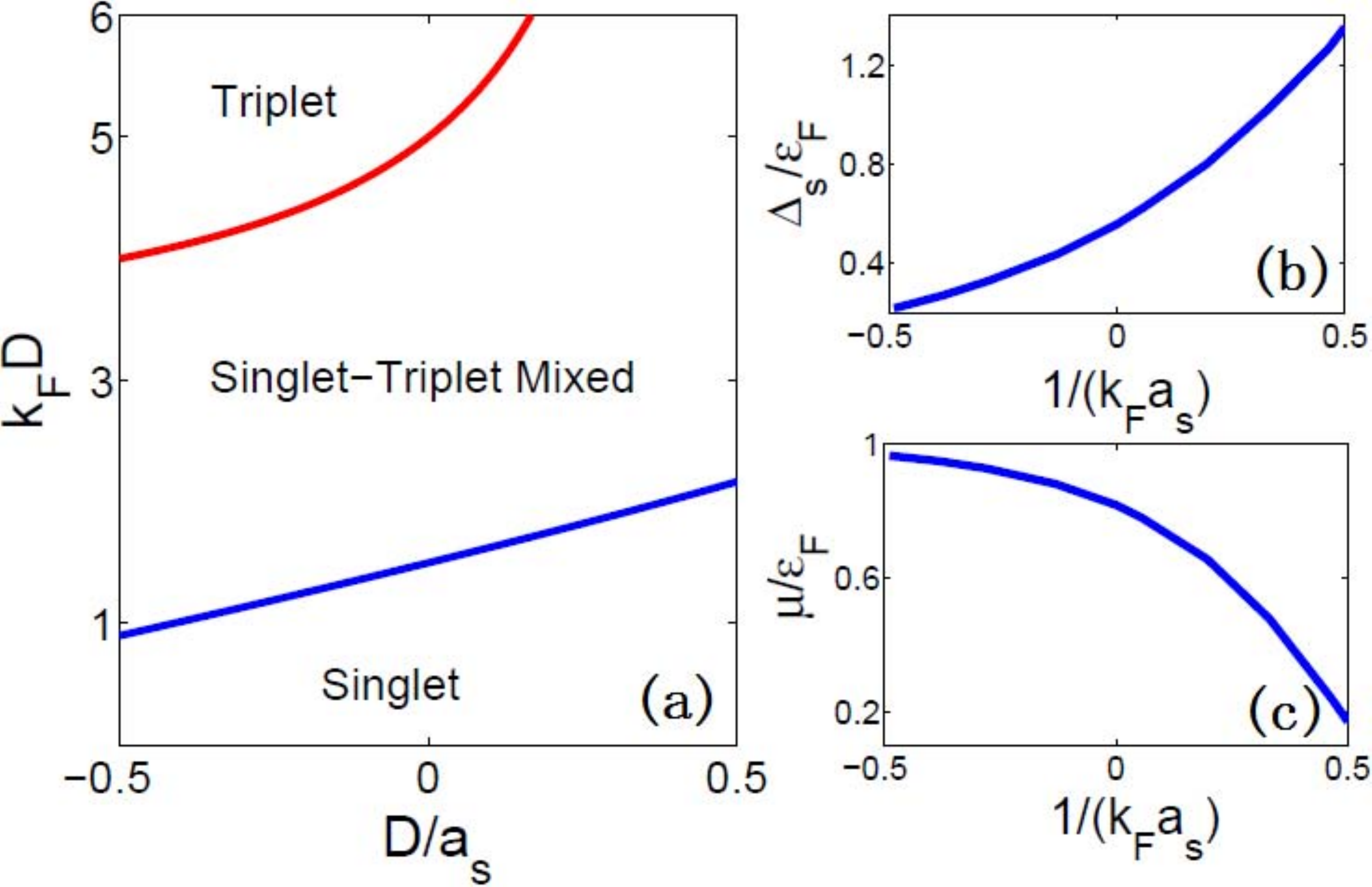}
\caption{(a) Schematic of the phase diagram for spin-$\frac{1}{2}$ Fermi gases of polar molecules across a DIIR. (b-c), In the singlet pairing phase, $\Delta/\mathcal{E}_{\text{F}}$ (b) and $\mu/\mathcal{E}_{\text{F}}$ (c) as a function of $1/k_{\text{F}}a_{\text{s}}$.  \label{pdg}}
\end{figure}

Finally we show in Fig. \ref{pdg} (b) and (c) that in the singlet phase, the system display standard behavior of BCS-BEC crossover, that is, the pairing gap continuously increases and the chemical potential continuously decreases as a function of $1/(k_{\text{F}}a_{\text{s}})$.

{\it (B) Large $k_{\text{F}}D$ Regime:} As $k_{\text{F}}D$ increases, we first find that in the singlet pairing phase, the $d$-wave component and higher partial wave component gradually increase, as shown in Fig. \ref{Deltal}(b-c). On the other hand, the energy for pure triplet pairing ($\Delta_{{\text{t}}{\bf k}}\neq 0$ and $\Delta_{{\text{s}}{\bf k}}=0$) gradually decreases, this is because the triplet pairing is dominated by $p$-wave as shown in Fig. \ref{Deltal}(c), and $a_{\text{p}}$ is proportional to $D$ as shown in Fig. \ref{as}(d). Therefore, as $k_{\text{F}}D$ increases, the energy of the triplet pairing phase will become lower than the single pairing phase. At DIIR, we find that this energy level crossing occurs around $k_FD\simeq 2$. Moreover, in the regime when singlet and triplet paring energy are close, we find a new superfluid state with nonzero $\Delta_{{\text{s}}{\bf k}}$ and $\Delta_{{\text{t}}{\bf k}}$ whose energy is lower than pure singlet and pure triplet paring, as shown in Fig. \ref{energy}(c-d). Since $\Delta_{{\text{s}}}$ and $\Delta_{{\text{t}}}$ have opposite parity under the time-reversal transformation, thus, time-reversal symmetry will be spontaneously broken when they coexist \cite{congjun}. As shown in Fig. \ref{energy}(c-d), the parameter window for mixed paring state decreases as one moves to the BCS side of the DIIR. Based on the energy comparison, a phase diagram is constructed as shown in Fig. \ref{pdg}(a), in which two second order phase boundaries separate singlet, mixed and triplet paring phases.

Hence, we have established results (A) and (B).

{\it Final Comments:} At last, let us make two comments. Firstly, we notice that the Fock exchange energy will deform the Fermi surface \cite{han,JN,Fradkin,Wu} and has influence on fermion pairing \cite{shitao,renyuan}. Such effect can be included by replacing the bare dispersion $\epsilon_{\mathbf{k}}$ with $\epsilon_{\mathbf{k}}+\Sigma_{\mathbf{k}}$ where $\Sigma_{\mathbf{k}}$ is the Fock self-energy given as
$\Sigma_{\mathbf{k}}=-\sum_{\mathbf{q}}V_{d}(\mathbf{k-q})n_{\mathbf{q}}/2$.
This equation for $\Sigma_{\mathbf{k}}$ needs to be solved together with Eqs. \ref{gap} self-consistently. We find that including this Fock exchange term does not lead to any qualitative change to the results discussed above. It only quantitatively shifts the phase boundary to lower value of $k_{\text{F}}D$, which implies that the deformation of Fermi surface seems to favor triplet pairing over singlet paring. At small $k_\text{F}D\lesssim 0.5$, even the quantitative correction to the ground state energy from the Fock term is negligibly small.

Secondly, we notice that there are several papers using $\Gamma_{\mathbf{k,k'}}=4\pi a_s/m+V_{\text{d}}(\mathbf{k-k'})$ to study fermion pairing problem \cite{congjun,shitao,renyuan}, where they assume $a_{\text{s}}$ is originated from the short-range potential, and $V_{\text{d}}$ is taken as the long-range dipolar potential.   Such a treatment is also inappropriate near a DIIR. This is because although the interaction potential is a combination of the short-range and the long-range part, the scattering amplitude can not be taken as a sum of two contributions unless one takes Born approximation. Furthermore, in this treatment, if $a_s\gg D$ the contribution from dipolar part completely vanishes for all range of $k_{\text{F}}D$. Therefore, it can not capture the finite range correction for small $k_{\text{F}}D$ and the phase transitions for large $k_{\text{F}}D$.

{\it Acknowledgements.} We thank Tin-Lun Ho, Georgy Shlyapnikov and Eugene Demler for helpful discussions. This work is supported by Tsinghua University Initiative Scientific Research Program. HZ is supported by NSFC under Grant No. 11004118 and No. 11174176, and NKBRSFC under Grant No. 2011CB921500. RQ is supported by NSFC under Grant No. 11104157.

{\it Note Added:} During writing this manuscript, we become aware of another paper where the DIIR is modeled in a completely different way \cite{Yi} for studying many-body physics. The focus of these two papers are also different. Both the finite range effect and the phase transition between singlet and triplet phases are not discussed there.


\begin{thebibliography}{99}
\bibitem{note}
The coupling between $s$ and $d$ wave is given by $D/r^3$, and the energy denominate is $2/r^2$, thus, the second order perturbation lowers the energy by $-D^2/r^4$.
\bibitem{SU2}
For ferromagnetic configuration, we have $\lambda_{1\mathbf{k}}^2-\lambda_{2\mathbf{k}}^2=2|\Delta_{t\mathbf{k}}|\sqrt{|\Delta_{s\mathbf{k}}|^2+|\Delta_{t\mathbf{k}}|^2}>0$ and thus the
pairing gap of different spin components are always different which will lead to spontaneous magnetization. For pure triplet pairing $\Delta_{s\mathbf{k}}=0$, one can prove rigorously that the condensation energy in ferromagnetic configuration is only one half of that in polar configuration and thus energetically unfavored.
\end{thebibliography}
\end{document}